\documentclass[conference]{IEEEtran}
\IEEEoverridecommandlockouts

\usepackage{cite}
\usepackage{amsmath,amssymb,amsfonts}
\usepackage{algorithmic}
\usepackage{graphicx}
\usepackage{textcomp}
\usepackage{xcolor}
\usepackage{microtype}
\usepackage{multirow}
\usepackage{booktabs}
\usepackage[normalem]{ulem}
\usepackage{xparse}
\usepackage{etoolbox}

\makeatletter
\patchcmd{\@makecaption}{\\}{:\enskip}{}{}
\makeatother

\usepackage{glossaries}

\newacronym{isac}{ISAC}{integrated sensing and communication}
\newacronym{rcs}{RCS}{radar cross-section}
\newacronym{uav}{UAV}{unmanned aerial vehicle}
\newacronym{pta}{PTA}{phase-time array}
\newacronym{pa}{PA}{phased array}
\newacronym{fd}{FD}{fully digital}
\newacronym{ttd}{TTD}{true time delay}
\newacronym{gdop}{GDOP}{geometric dilution of precision}
\newacronym{wls}{WLS}{weighted least squares}
\newacronym{mlp}{MLP}{multilayer perceptron}
\newacronym{cnn}{CNN}{convolutional neural network}
\newacronym{ps}{PS}{phase shifter}
\newacronym{ofdm}{OFDM}{orthogonal frequency division multiplexing}
\newacronym{fft}{FFT}{fast Fourier transform}
\newacronym{ifft}{IFFT}{inverse fast Fourier transform}
\newacronym{bs}{BS}{base station}
\newacronym{ula}{ULA}{uniform linear array}
\newacronym{rf}{RF}{radio frequency}
\newacronym{tx}{Tx}{transmitter}
\newacronym{rx}{Rx}{receiver}
\newacronym{aoa}{AoA}{angle of arrival}
\newacronym{aod}{AoD}{angle of departure}
\newacronym{roi}{ROI}{region of interest}
\newacronym{amcf-zci}{AMCF-ZCI}{alternating minimization for constrained Frobenius-norm with Zadoff-Chu sequence initialization}
\newacronym{fim}{FIM}{Fisher information matrix}
\newacronym{rmse}{RMSE}{root mean square error}
\newacronym{cdf}{CDF}{cumulative distribution function}
\newacronym{snr}{SNR}{signal-to-noise ratio}
\newacronym{ls}{LS}{least squares}
\newacronym{isd}{ISD}{inter-site distance}

\def\BibTeX{{\rm B\kern-.05em{\sc i\kern-.025em b}\kern-.08em
    T\kern-.1667em\lower.7ex\hbox{E}\kern-.125emX}}
\begin{document}

\newcommand{\jmo}[1]{\textcolor{red}{XXX jmo: #1}}

\title{Phase-Time Array Enabled Multistatic Sensing with Multi-Level Fusion for UAV Localization
}

\author{\IEEEauthorblockN{Ming Gao, Jianhua Mo, Meixia Tao}
    \IEEEauthorblockA{School of Information Science and Electronic Engineering,
        Shanghai Jiao Tong University, Shanghai, China \\
        Emails: \{gaoming25, mjh, mxtao\}@sjtu.edu.cn}
    \thanks{This work is supported by the National Natural Science Foundation of China (NSFC) under Grant 62125108 and by the Science and Technology Commission Foundation of Shanghai under Grant 25DP1501200.}
}

\maketitle

\begin{abstract}
    In multistatic collaborative sensing, existing parameter-level fusion methods suffer from severe error amplification in geometrically unfavorable regions. Conventional receiver architectures present an additional challenge: analog phased arrays incur substantial beam sweeping overhead, while fully digital arrays demand prohibitive hardware costs. This paper proposes a multistatic sensing framework based on a phase-time array (PTA), whose rainbow beamforming maps spatial directions to OFDM subcarriers, achieving wide-angle coverage with a single RF chain and a favorable accuracy--complexity trade-off relative to both phased array and fully digital receivers. To suppress geometry-induced error hotspots, two parameter-level fusion schemes are developed: a GDOP-weighted analytical estimator (GDOP-WLS) and a lightweight multilayer perceptron (PF-MLP). An end-to-end signal-level convolutional neural network (SF-CNN) is further designed to directly regress target coordinates from raw received signals, bypassing cascaded parameter estimation. Simulation results show that the parameter-level schemes effectively mitigate localized error amplification with minimal computational overhead, while SF-CNN attains sub-meter positioning accuracy, at the cost of higher model complexity that can be offset by GPU acceleration. Power-matched training further extends both schemes' accuracy across a wide power~range.
\end{abstract}

\glsresetall


\section{Introduction}

Driven by the rapid growth of the low-altitude economy, \glspl{uav} have been widely deployed in urban environments for applications such as logistics delivery, infrastructure inspection, and emergency response\cite{jiang2025integrated}. However, this widespread accessibility also increases the risk of unauthorized or malicious flights, making the detection and accurate localization of these non-cooperative \glspl{uav} a major challenge for airspace management\cite{bi2025opportunities}. \Gls{isac} networks, which leverage existing communication \glspl{bs} and spectrum resources to simultaneously perform data transmission and environmental sensing, offer a cost-effective platform for large-scale \gls{uav} surveillance without requiring dedicated radar infrastructure\cite{liu2022integrated}. Nevertheless, monostatic \gls{isac} sensing inherently suffers from full-duplex self-interference and fluctuating target \gls{rcs}, which can severely degrade detection reliability\cite{lu2023degrees,barneto2019full}. Multistatic collaborative sensing provides an attractive alternative by exploiting spatial diversity across distributed nodes, thereby improving localization robustness~\cite{xie2023collaborative,liu2024cooperative}.

Realizing multistatic collaborative sensing in practice, however, presents significant hardware and efficiency challenges, primarily due to the limitations of conventional array architectures. Specifically, traditional analog \glspl{pa} rely on time-division beam sweeping; coordinating sequential sweeps across multiple distributed nodes incurs substantial synchronization overhead and limits the update rate for dynamic targets. Conversely, \gls{fd} arrays avoid sweeping by forming multiple simultaneous beams in the digital domain, but each beam requires a dedicated \gls{rf} chain, resulting in large hardware cost and power consumption that hinder large-scale deployment.

An alternative is the \gls{pta} architecture, which combines \glspl{ps} with \gls{ttd} units\cite{ratnam2022joint,nam2025joint,cai2026hybrid}. Through proper configuration of the phase shifts and time delays, the \gls{pta} can steer different \gls{ofdm} subcarriers toward distinct spatial directions, forming frequency-dependent rainbow beams\cite{luo2024yolo,liang2025cfarnet,cai2026spot} that cover a wide angular sector with a single \gls{rf} chain. This allows each receiving node in a multistatic network to detect targets from unknown directions without beam sweeping\cite{cai2026spot}.

While the \gls{pta} architecture addresses the hardware and efficiency challenges of data acquisition, the final localization accuracy of a multistatic collaborative system critically depends on how multistatic measurements are fused\cite{wei2024symbollevel,wang2025cooperative,yan2026asynchronous}. A common approach is parameter-level fusion \cite{liu2024cooperative}, where each \gls{rx} first independently estimates local parameters such as angle and range, and a central processor then combines these estimates to recover the target position. However, such schemes are highly sensitive to the spatial geometry between the \glspl{bs} and the target, a vulnerability commonly quantified by the \gls{gdop}\cite{kanhere2021target}. When the target falls into geometrically unfavorable configurations, even small estimation errors can be greatly amplified during fusion, creating geometry-induced error hotspots\cite{nguyen2016optimal, fatima2024optimal}. Fusing directly at the signal level can circumvent such cascaded estimation errors, at the cost of increased model complexity.

To address these issues, we propose a multistatic sensing framework based on the \gls{pta} architecture and advanced fusion strategies. The main contributions are summarized as follows:

\begin{itemize}
    \item We introduce the \gls{pta} architecture into multistatic collaborative sensing. Unlike \gls{pa} receivers that resolve range only, the \gls{pta} recovers both angle and range from a single \gls{rf} chain, enabling the downstream geometry-aware fusion at an order-of-magnitude lower latency than \gls{fd}.
    \item \looseness=-1 We propose two parameter-level fusion schemes to suppress geometry-induced error hotspots: a least squares estimator with GDOP-weighted initialization (GDOP-WLS) and a parameter-fusion multilayer perceptron (PF-MLP), which reduce the positioning \gls{rmse} from 11.71~m to 1.75~m and 1.66~m, respectively, with negligible latency~overhead.
    \item We design an end-to-end signal-fusion convolutional neural network (SF-CNN) that directly regresses target coordinates from the received signals of all nodes, bypassing cascaded parameter estimation and achieving a positioning \gls{rmse} of 0.44~m. Power-matched training further extends both neural-network-based schemes' accuracy across a wide transmit power range.
\end{itemize}

\section{System Model}
\label{sec:system_model}

\begin{figure}[!t]
    \centering
    \includegraphics[width=0.85\linewidth]{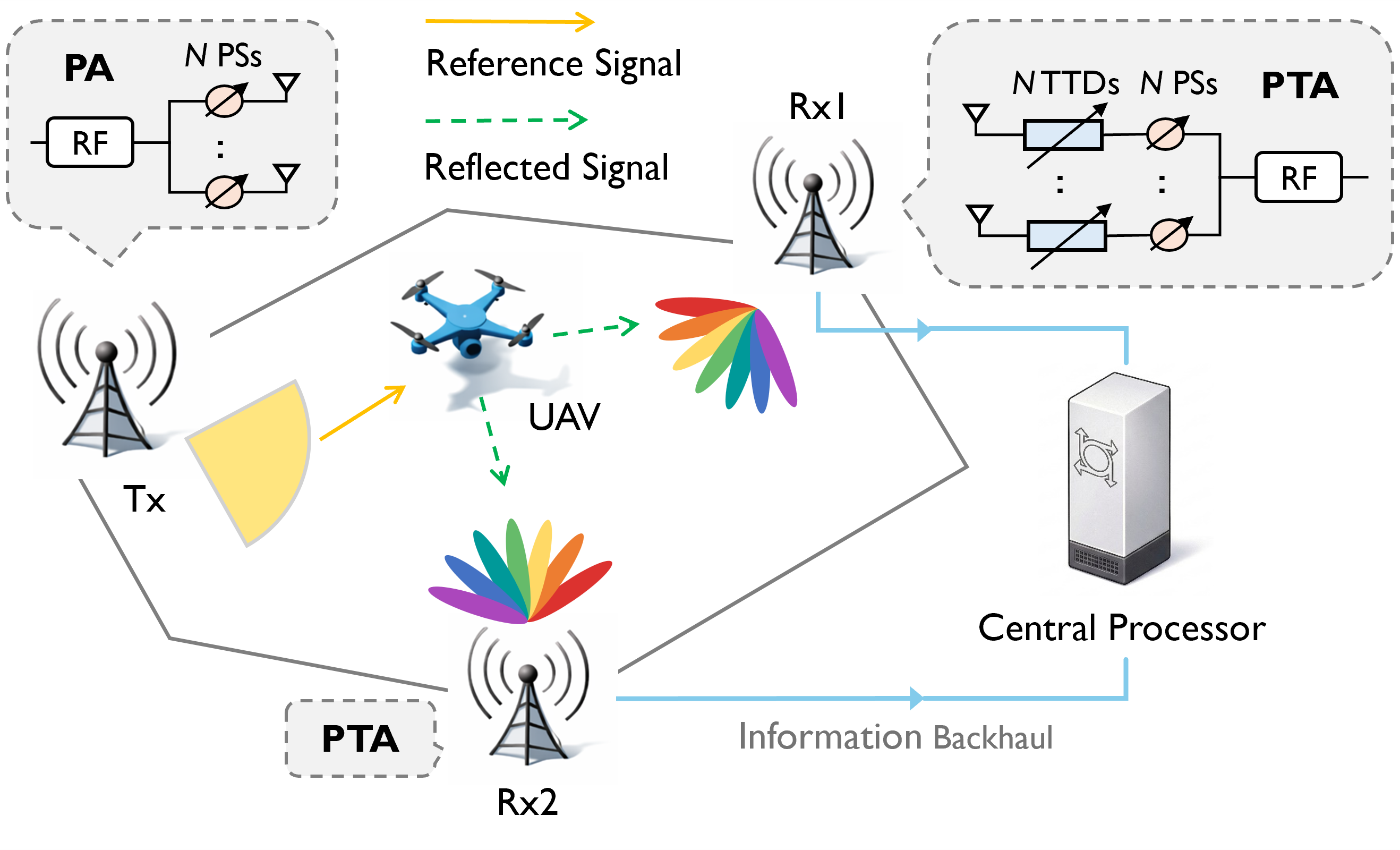}
    \vspace{-0.5em}
    \caption{System model.}
    \label{fig:system_model}
    \vspace{-1.5em}
\end{figure}

As illustrated in Fig.~\ref{fig:system_model}, we consider a multistatic \gls{isac} system for collaborative \gls{uav} localization, where one \gls{bs} acts as the \gls{tx} and two geographically separated base stations serve as receivers (\gls{rx}1, \gls{rx}2). Specifically, these nodes are deployed with an \gls{isd} $L$ at coordinates $\mathbf{p}_{tx}$, $\mathbf{p}_{rx,1}$, and $\mathbf{p}_{rx,2}$, respectively, with distinct boresight orientations. The target \gls{uav} is uniformly distributed over a hexagonal \gls{roi} and is assumed to share the same altitude as the \glspl{bs} to simplify the analysis. Each \gls{bs} is equipped with an $N$-element \gls{ula} with inter-element spacing $d$. The \gls{tx} employs a conventional \gls{pa} architecture to form a fixed wide beam at low complexity, while each receiving array adopts a \gls{pta} architecture driven by a single \gls{rf} chain. The system operates on a single \gls{ofdm} symbol with $N_c$ subcarriers, subcarrier spacing $\Delta f$, carrier frequency $f_c$, and wavelength $\lambda$.

In each bistatic link, the transmitted signal propagates to the target and is scattered back to the corresponding \gls{rx}. For a target located at position $\mathbf{p} = (x,y)$, the \gls{tx}-to-target distance $R_{tx}$ and the target-to-\gls{rx}$i$ distance $R_{rx,i}$ are defined respectively as
\begin{small}
    \begin{equation}
        R_{tx} = \sqrt{(x-x_{tx})^2+(y-y_{tx})^2},
    \end{equation}
    \begin{equation}
        R_{rx,i} = \sqrt{(x-x_{rx,i})^2+(y-y_{rx,i})^2}, \quad i=1,2,
    \end{equation}
\end{small}%
where $(x_{tx}, y_{tx})$ and $(x_{rx,i}, y_{rx,i})$ denote the Cartesian components of the position vectors $\mathbf{p}_{tx}$ and $\mathbf{p}_{rx,i}$, respectively.

\subsection{PTA-based Rainbow Beamforming}
\label{subsec:rainbow_bf}

Within the receiving \gls{pta} architecture, the $n$-th antenna element is equipped with a programmable \gls{ps} of value $\phi_n$ and a \gls{ttd} unit with delay $t_n$. Let $f_m = f_0 + m\Delta f$ denote the absolute frequency of the $m$-th \gls{ofdm} subcarrier, where $f_0$ is the starting subcarrier frequency. The $n$-th element of the beamforming weight vector $\mathbf{w}_m \in \mathbb{C}^{N \times 1}$ for the $m$-th subcarrier is given by
\begin{equation}
    [\mathbf{w}_m]_n
    = \frac{1}{\sqrt{N}}
    \exp\Bigl(-j2\pi\phi_n\Bigr)\,
    \exp\Bigl(-j2\pi(f_m - f_0)\, t_n\Bigr).
    \label{eq:weight_vector}
\end{equation}

To steer the main lobe from $\theta_{\mathrm{start}}$ at the starting subcarrier frequency $f_0$ to $\theta_{\mathrm{end}}$ at the highest frequency $f_{\mathrm{high}} = f_0 + W$ with $W \triangleq (N_c - 1)\Delta f$, the phase shift and time delay are configured as \cite{luo2024yolo}
\begin{equation}
    \phi_n = -\frac{f_0 n d\sin\theta_{\mathrm{start}}}{c},
\end{equation}
\begin{equation}
    t_n = \frac{nd}{Wc}\bigl(f_0\sin\theta_{\mathrm{start}} - f_{\mathrm{high}}\sin\theta_{\mathrm{end}}\bigr),
\end{equation}
where $c$ is the speed of light. This establishes a one-to-one mapping between the \gls{ofdm} subcarriers and spatial directions across $[\theta_{\mathrm{start}}, \theta_{\mathrm{end}}]$.

\subsection{Received Signal Model}
\label{subsec:signal_model}

Based on the wideband bistatic geometry, the frequency-domain channel matrix $\mathbf{H}_{m}^{(i)}\in\mathbb{C}^{N\times N}$ for the $i$-th link ($i=1,2$) on the $m$-th subcarrier is modeled as
\begin{equation}
    \mathbf{H}_{m}^{(i)} = \beta_i\, e^{-j2\pi f_m\tau_i}\, \mathbf{a}_{r,i}(f_m,\theta_i)\,\mathbf{a}_t^T(f_m,\phi),
    \label{eq:channel_matrix}
\end{equation}
where $\theta_i$ and $\phi$ denote the \gls{aoa} at the $i$-th \gls{rx} and the \gls{aod} from the \gls{tx}, respectively. The bistatic propagation delay for link $i$ is given by $\tau_i = (R_{tx}+R_{rx,i})/c$, and the path gain coefficient is modeled as
\begin{equation}
    \beta_i = \sqrt{\frac{\lambda^2\sigma_{\text{rcs}}}{(4\pi)^3 R_{tx}^2\, R_{rx,i}^2}},
    \label{eq:beta_i}
\end{equation}
where $\sigma_{\text{rcs}}$ represents the \gls{rcs} of the target. The transmit and receive steering vectors are frequency-dependent and given respectively by
\begin{equation}
    [\mathbf{a}_t(f_m,\phi)]_n = \exp\Bigl(j2\pi f_m \frac{nd\sin\phi}{c}\Bigr),
\end{equation}
\begin{equation}
    [\mathbf{a}_{r,i}(f_m,\theta_i)]_n = \exp\Bigl(j2\pi f_m \frac{nd\sin\theta_i}{c}\Bigr).
\end{equation}

To ensure uniform illumination over the entire \gls{roi}, the \gls{tx} employs a spatial beamformer $\mathbf{v}\in\mathbb{C}^{N\times 1}$ synthesized via the \gls{amcf-zci} algorithm~\cite{qi2020hierarchical}. Let $P_{tx}$ denote the total transmit power and $s_{m}$ be the transmitted pilot symbol, normalized such that $\mathbb{E}[|s_{m}|^2] = 1/N_c$. The received baseband signal at the $i$-th \gls{rx} on the $m$-th subcarrier is given by
\begin{equation}
    y_{m}^{(i)} = \sqrt{P_{tx}}\,\mathbf{w}_m^H\mathbf{H}_{m}^{(i)}\mathbf{v}s_{m} + n_{m}^{(i)},
    \label{eq:received_signal_scalar}
\end{equation}
where $n_{m}^{(i)} \sim \mathcal{CN}(0,\, N_0 \Delta f)$ denotes the additive white Gaussian noise, with $N_0$ the noise power spectral density. The direct-path signal from the \gls{tx} to each \gls{rx} is assumed to be suppressed beforehand and is thus not considered in~\eqref{eq:received_signal_scalar}.

\section{Proposed PTA-Enabled Multistatic Collaborative Localization Framework}
\label{sec:algorithm}

As illustrated in Fig.~\ref{fig:framework}, the proposed framework comprises a parameter-level paradigm, which performs bistatic parameter estimation (Section~\ref{subsec:estimation}) followed by parameter-level fusion (Section~\ref{subsec:fusion}), and a signal-level paradigm (Section~\ref{subsec:sfcnn}).

\subsection{Bistatic Parameter Estimation}
\label{subsec:estimation}

The parameter estimation pipeline operates in two stages: the target \gls{aoa} is first extracted via peak subcarrier detection by exploiting the spatial-spectral coupling of rainbow beamforming, and the bistatic propagation distance is subsequently estimated by applying a zero-padded \gls{ifft} to a sub-band of the received signal centered around the detected peak subcarrier.

\subsubsection{Angle Estimation via Peak Subcarrier Detection}

Due to the spatial-spectral coupling of rainbow beamforming, the received power is maximized at the subcarrier corresponding to the target \gls{aoa}. Consequently, the spatial direction can be efficiently extracted by locating the peak-power index of the received signal $y_{m}^{(i)}$:
\begin{equation}
    m_i^* = \operatorname*{arg\,max}_{m}\, \bigl|y_{m}^{(i)}\bigr|^2.
    \label{eq:peak_subcarrier}
\end{equation}

Let $f_{m_i^*} = f_0 + m_i^*\Delta f$ be the corresponding absolute frequency.
Inverting the \gls{pta} frequency-to-angle mapping defined in Section~\ref{subsec:rainbow_bf}~\cite{luo2024yolo} yields the closed-form \gls{aoa} estimate:
\begin{small}
    \begin{equation}
        \sin\hat{\theta}_i = \frac{f_0 \bigl(f_{\mathrm{high}} - f_{m_i^*}\bigr)}{W f_{m_i^*}}\sin\theta_{\mathrm{start}} + \frac{f_{\mathrm{high}}\bigl(f_{m_i^*} - f_0\bigr)}{W f_{m_i^*}}\sin\theta_{\mathrm{end}}.
        \label{eq:freq_angle_map}
    \end{equation}
\end{small}
\subsubsection{Distance Estimation via Sub-band \gls{ifft}}

\looseness=-1
With $\hat{\theta}_i$ determined, the bistatic propagation distance is estimated from a sub-band $\mathcal{M}_{\mathrm{sub}}$ of $M_{\mathrm{sub}}$ subcarriers centered at the peak index $m_i^*$. The sub-band width $M_{\mathrm{sub}}$ is set equal to the 3~dB main-lobe width of the subcarrier power spectrum $|y_m^{(i)}|^2$, thereby confining the processing window to the high-\gls{snr} region of the rainbow beam response while suppressing out-of-lobe noise and beam-squint-induced steering mismatch.

The sub-band signal vector $\mathbf{y}_{\mathrm{sub}}^{(i)} \in \mathbb{C}^{M_{\mathrm{sub}} \times 1}$ is zero-padded to length $N_{\mathrm{IFFT}} \geq M_{\mathrm{sub}}$ and transformed to the delay domain via the \gls{ifft}:
\begin{equation}
    \mathbf{r}^{(i)} = \mathbf{F}_{N_{\mathrm{IFFT}}}^{H}\, \tilde{\mathbf{y}}_{\mathrm{sub}}^{(i)},
    \label{eq:ifft_range}
\end{equation}
where $\tilde{\mathbf{y}}_{\mathrm{sub}}^{(i)} \in \mathbb{C}^{N_{\mathrm{IFFT}} \times 1}$ is the zero-padded signal vector and $\mathbf{F}_{N_{\mathrm{IFFT}}}$ denotes the $N_{\mathrm{IFFT}}$-point unitary DFT matrix with entries $[\mathbf{F}_{N_{\mathrm{IFFT}}}]_{k,q} = N_{\mathrm{IFFT}}^{-1/2}\, e^{-j2\pi kq/N_{\mathrm{IFFT}}}$. Each delay bin $k$ maps to a candidate bistatic distance $r_k = kc/(N_{\mathrm{IFFT}} \Delta f)$, and the estimated distance is obtained as
\begin{equation}
    \hat{d}_i = \operatorname*{arg\,max}_{r_k}\, \bigl|[\mathbf{r}^{(i)}]_k\bigr|^2.
    \label{eq:ifft_distance}
\end{equation}

\subsection{Parameter-Level Fusion Strategies}
\label{subsec:fusion}

In the parameter-level fusion stage, the central processor fuses the intermediate parameter estimates $\{\hat{\theta}_i, \hat{d}_i\}_{i=1}^{2}$ to recover the target position.

\subsubsection{Measurement Model}

Given a candidate target position $\mathbf{p} = (x,y)$, the corresponding \gls{tx}-to-target and target-to-\gls{rx}$i$ distances are $R_{tx}$ and $R_{rx,i}$, as defined in Section~\ref{sec:system_model}. The noiseless measurement functions for link $i$ are then
\begin{align}
    d_i(\mathbf{p})      & = R_{tx} + R_{rx,i}, \label{eq:meas_d}                                                      \\
    \theta_i(\mathbf{p}) & = \mathop{\mathrm{atan2}}\!\left(y - y_{rx,i},\, x - x_{rx,i}\right), \label{eq:meas_theta}
\end{align}
where $d_i(\mathbf{p})$ is the propagation distance, $\theta_i(\mathbf{p})$ represents the \gls{aoa} in the global coordinate system, and $\mathrm{atan2}$ denotes the four-quadrant inverse tangent function. The observed quantities $\hat{d}_i$ and $\hat{\theta}_i$ are modeled as $d_i(\mathbf{p})$ and $\theta_i(\mathbf{p})$ corrupted by zero-mean errors with standard deviations $\sigma_{d,i}$ and $\sigma_{\theta,i}$, respectively.

\subsubsection{GDOP-weighted Initialization}

To prevent the nonlinear \gls{wls} solver from converging to local minima, a reliable initialization is essential. For each link $i$, a closed-form coarse position estimate $\hat{\mathbf{p}}_{\mathrm{geo}}^{(i)}$ is calculated by finding the geometric intersection of the bistatic ellipse defined by $\hat{d}_i$ and the \gls{aoa} ray defined by $\hat{\theta}_i$:
\begin{equation}
    \hat{\mathbf{p}}_{\mathrm{geo}}^{(i)} = \mathbf{p}_{rx,i} + \frac{\hat{d}_i^2 - \|\mathbf{p}_{rx,i} - \mathbf{p}_{tx}\|^2}{2\bigl[\hat{d}_i + (\mathbf{p}_{rx,i} - \mathbf{p}_{tx})^T \mathbf{u}_i\bigr]} \mathbf{u}_i,
    \label{eq:geo_init}
\end{equation}
where $\mathbf{u}_i = [\cos\hat{\theta}_i, \sin\hat{\theta}_i]^T$ is the directional unit vector.

The positioning reliability of each link depends on the local geometric configuration, requiring the per-link \gls{gdop} to be quantified.
Let $\mathbf{z}_i = [d_i,\, \theta_i]^T$ denote the measurement vector. Taking the partial derivatives of $d_i(\mathbf{p})$ and $\theta_i(\mathbf{p})$ from \eqref{eq:meas_d} and \eqref{eq:meas_theta} yields the geometric Jacobian matrix $\mathbf{J}_i = \partial \mathbf{z}_i / \partial \mathbf{p} \in \mathbb{R}^{2 \times 2}$ evaluated at $\hat{\mathbf{p}}_{\mathrm{geo}}^{(i)}$:
\begin{equation}
    \mathbf{J}_i =
    \begin{bmatrix}
        \dfrac{x - x_{tx}}{R_{tx}} + \dfrac{x - x_{rx,i}}{R_{rx,i}} & \dfrac{y - y_{tx}}{R_{tx}} + \dfrac{y - y_{rx,i}}{R_{rx,i}} \\[14pt]
        -\dfrac{y - y_{rx,i}}{R_{rx,i}^2}                           & \dfrac{x - x_{rx,i}}{R_{rx,i}^2}
    \end{bmatrix}\!.
    \label{eq:jacobian}
\end{equation}
Under the small-error assumption, the total differential of the measurement equations reduces to the linear relationship $d\mathbf{z}_i = \mathbf{J}_i \, d\mathbf{p}$. Consequently, the target position estimation error can be expressed as
\begin{equation}
    d\mathbf{p} = \left(\mathbf{J}_i^T \mathbf{\Sigma}_i^{-1} \mathbf{J}_i\right)^{-1} \mathbf{J}_i^T \mathbf{\Sigma}_i^{-1} \, d\mathbf{z}_i,
\end{equation}
where $\mathbf{\Sigma}_i = \mathrm{diag}(\sigma_{d,i}^2,\, \sigma_{\theta,i}^2)$ is the measurement error covariance matrix. Here $\sigma_{d,i}$ and $\sigma_{\theta,i}$ denote the estimation error standard deviations of the bistatic distance and \gls{aoa} for the $i$-th link, which are calibrated at each transmit power by Monte Carlo simulations of the parameter estimator described in Section~\ref{subsec:estimation} and kept fixed for all target positions. Under zero-mean uncorrelated measurement errors, the position error covariance reduces to $\mathbf{P}_i = (\mathbf{J}_i^T \mathbf{\Sigma}_i^{-1} \mathbf{J}_i)^{-1}$, whose trace equals the mean squared position error. The per-link \gls{gdop} is accordingly defined as
\begin{equation}
    \mathrm{GDOP}_i = \sqrt{\mathrm{trace}(\mathbf{P}_i)}.
    \label{eq:gdop}
\end{equation}
The \gls{gdop}-weighted initialization point is then established as
\begin{equation}
    \mathbf{p}_0 = \frac{w_1\,\hat{\mathbf{p}}_{\mathrm{geo}}^{(1)} + w_2\,\hat{\mathbf{p}}_{\mathrm{geo}}^{(2)}}{w_1 + w_2},
    \label{eq:weight_init}
\end{equation}
where $w_i = 1 / \mathrm{\gls{gdop}}_i$ denotes the initialization weight of the $i$-th sensing link.

\subsubsection{GDOP-WLS}

In the GDOP-WLS scheme, the final 2D position estimate is obtained by solving
\begin{equation}
    \hat{\mathbf{p}} = \operatorname*{arg\,min}_{\mathbf{p}} \sum_{i=1}^{2} \biggl[ \frac{\bigl(\hat{d}_i - d_i(\mathbf{p})\bigr)^2}{\sigma_{d,i}^2} + \frac{\bigl(\hat{\theta}_i \ominus \theta_i(\mathbf{p})\bigr)^2}{\sigma_{\theta,i}^2} \biggr]\!,
    \label{eq:wls_cost}
\end{equation}
where $\hat{\theta}_i \ominus \theta_i(\mathbf{p})$ is the wrapped angular difference in $(-\pi,\pi]$.
Normalization by the variances $\sigma_{d,i}^2$ and $\sigma_{\theta,i}^2$ makes the distance and angle terms dimensionless and mutually comparable.
The problem~\eqref{eq:wls_cost} is solved via the Levenberg--Marquardt algorithm initialized at $\mathbf{p}_0$.

\subsubsection{PF-MLP}

The PF-MLP scheme replaces the analytical optimization in GDOP-WLS with a compact \gls{mlp} that operates on the same intermediate parameter estimates $\{\hat{\theta}_i, \hat{d}_i\}_{i=1}^{2}$, directly regressing the 2D target coordinates. The network architecture is detailed in~Fig.~\ref{fig:framework}.

\subsection{Signal-Level Fusion Strategy}
\label{subsec:sfcnn}

\begin{figure}[!t]
    \centering
    \includegraphics[width=\columnwidth]{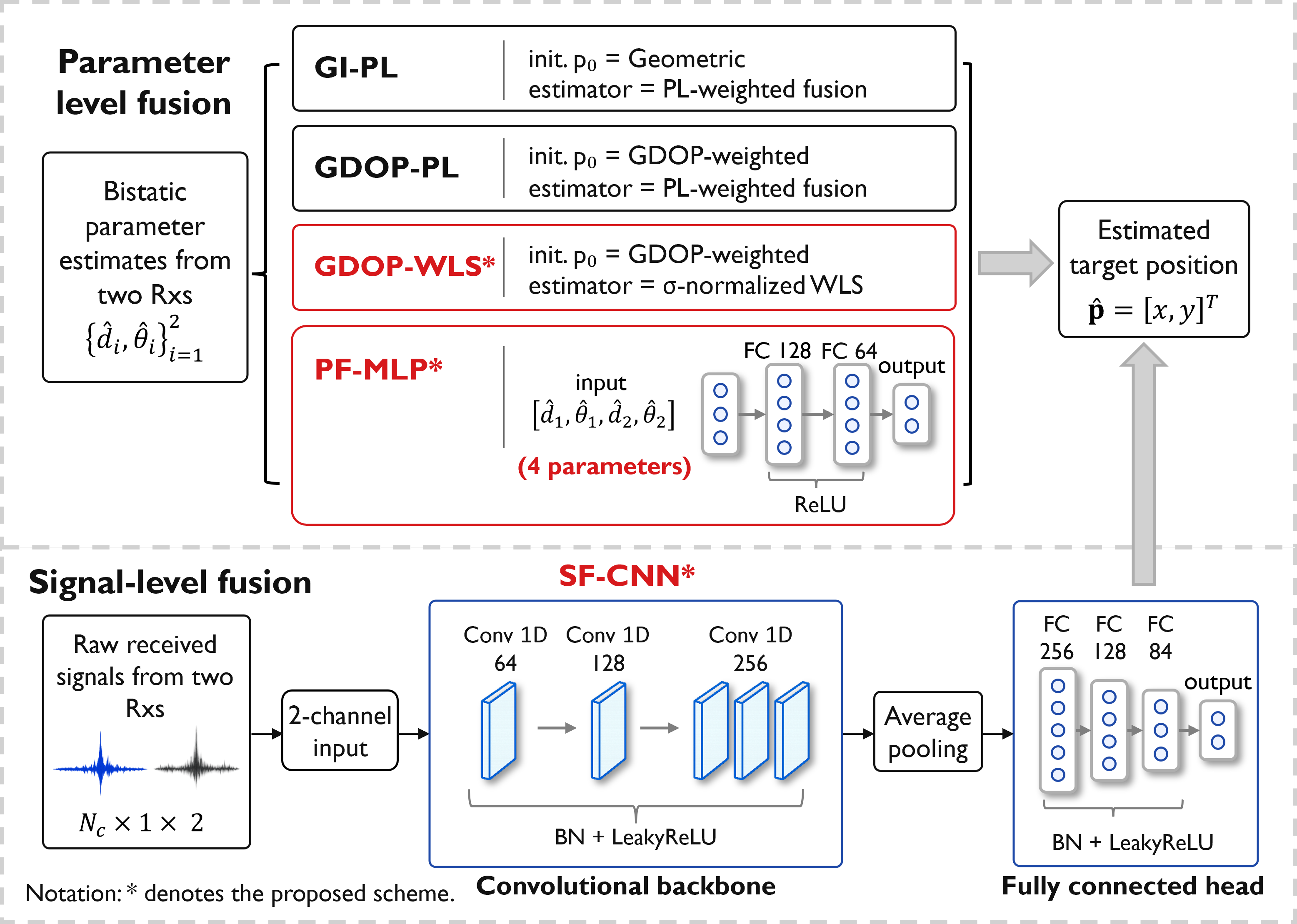}
    \vspace{-1.5em}
    \caption{Overview of the collaborative localization framework.}
    \label{fig:framework}
    \vspace{-1.5em}
\end{figure}

The SF-CNN scheme directly regresses the target coordinates from the received signal amplitudes $|y_m^{(i)}|$ of all distributed receivers, thereby avoiding cascaded parameter estimation errors. As shown in Fig.~\ref{fig:framework}, the network processes the two-channel \gls{rx} signals through a five-layer 1D convolutional backbone followed by average pooling and a fully connected regression head.

\section{Simulation Results}
\label{sec:simulation}

\begin{figure*}[!t]
    \centering
    \includegraphics[width=0.92\linewidth]{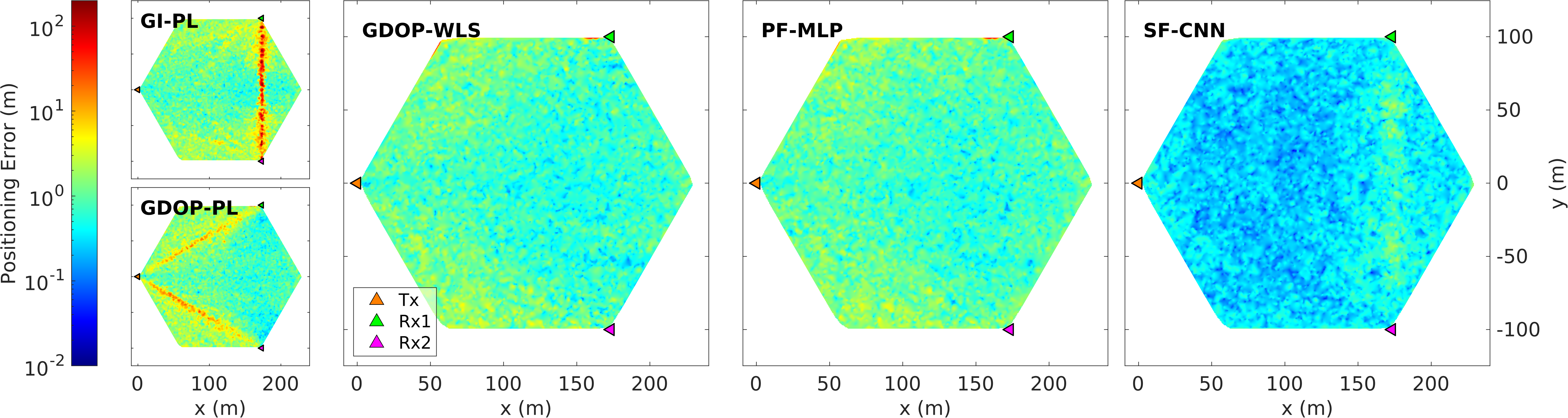}
    \vspace{-0.5em}
    \caption{Spatial distribution of positioning error for different fusion schemes.}
    \label{fig:heatmap}
    \vspace{-1.5em}
\end{figure*}

\begin{figure}[!t]
    \centering
    \includegraphics[width=0.8\linewidth]{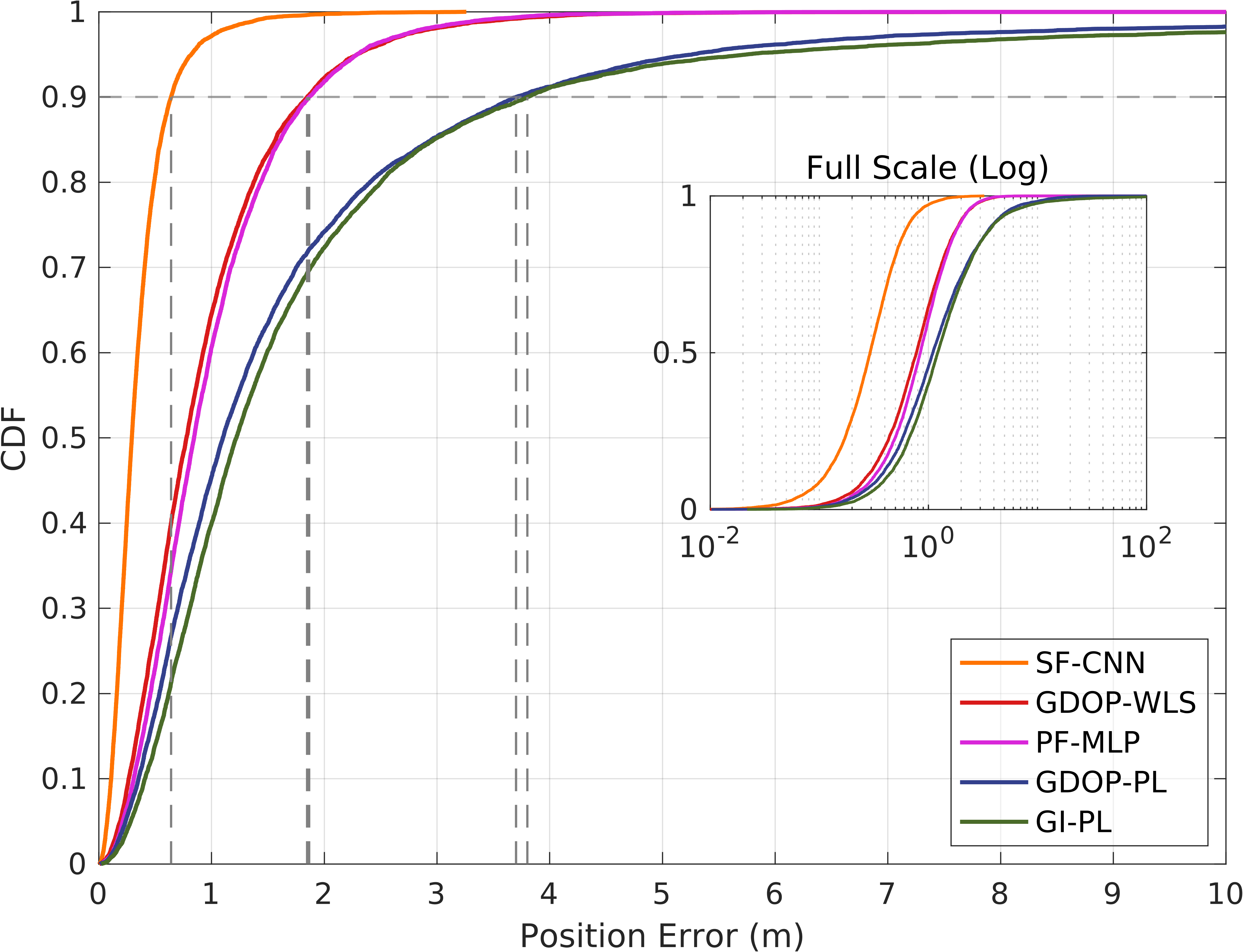}
    \vspace{-0.5em}
    \caption{Cumulative distribution function (CDF) of position error.}
    \label{fig:cdf}
    \vspace{-1em}
\end{figure}

\begin{figure}[!t]
    \centering
    \includegraphics[width=0.8\linewidth]{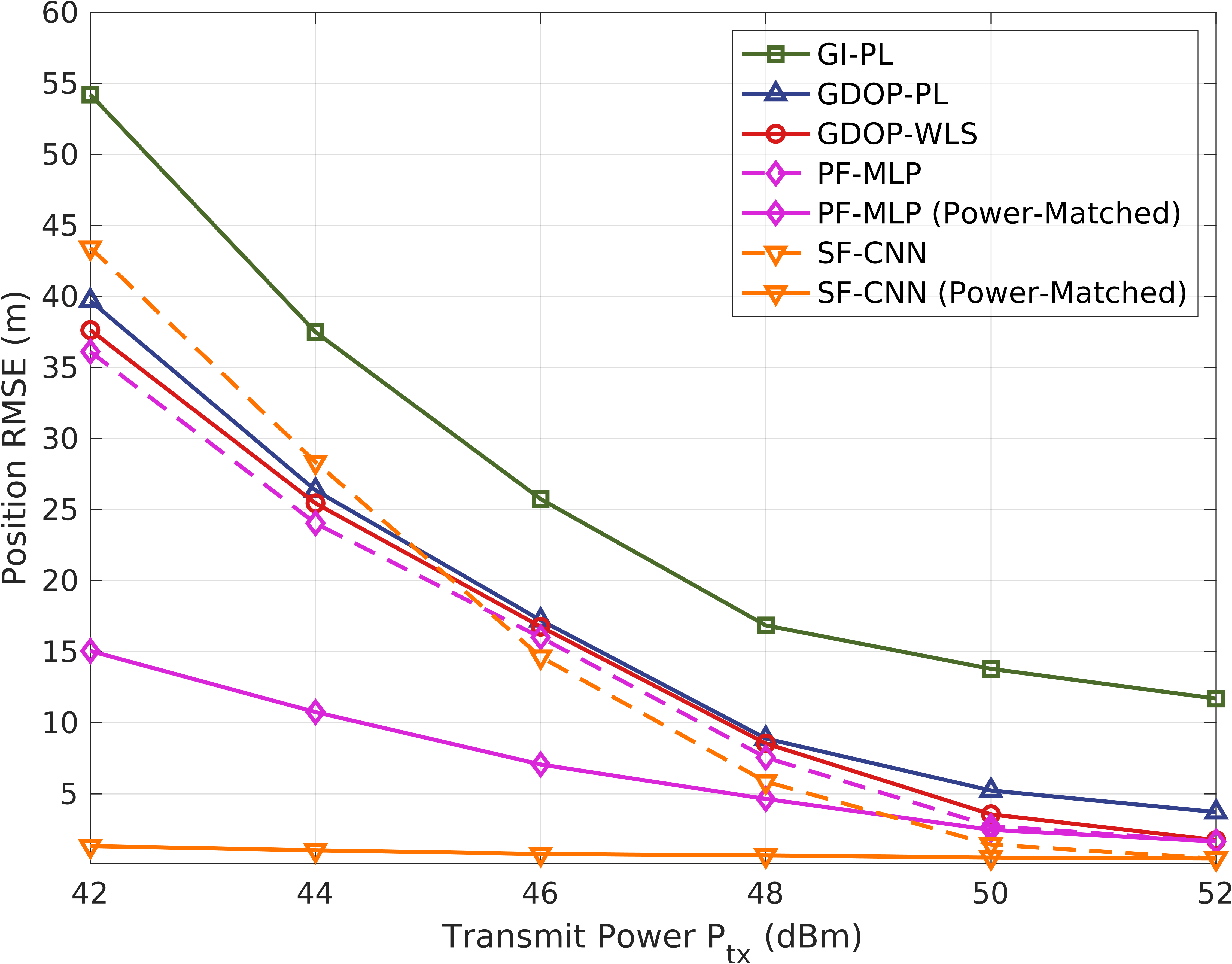}
    \vspace{-0.5em}
    \caption{Position RMSE versus transmit power $P_{tx}$.}
    \label{fig:rmse_vs_ptx}
    \vspace{-1.5em}
\end{figure}

\subsection{Simulation Setup}
\label{subsec:sim_setup}

Table~\ref{tab:sim_params} lists the default system parameters. PF-MLP and SF-CNN are trained on $100{,}000$ samples with an 8:1:1 training/validation/test split using the AdamW optimizer with MSE loss. The receiver architecture comparison (Section~\ref{subsec:rx_architecture}) and the fusion performance evaluation (Section~\ref{subsec:performance}) are conducted over the same $10{,}000$ independent Monte Carlo trials. All runtimes are measured on an NVIDIA RTX 5880 GPU and dual AMD EPYC 7Y43 CPUs. Excluding \gls{rf} acquisition, the reported latency comprises bistatic parameter estimation and fusion for the parameter-level schemes, and direct inference for SF-CNN.

\begin{table}[!t]
    \centering
    \caption{Simulation Parameters}
    \label{tab:sim_params}
    \vspace{-0.5em}
    \resizebox{0.99\columnwidth}{!}{%
        \begin{tabular}{l l}
            \hline
            \textbf{Parameter}                                                      & \textbf{Value}                                     \\
            \hline
            \gls{isd} $L$                                                           & 200 m                                              \\
            \gls{tx}/\gls{rx}1/\gls{rx}2 coordinates                                & $(0,0)$, $(\sqrt{3}L/2,L/2)$, $(\sqrt{3}L/2,-L/2)$ \\
            \gls{tx}/\gls{rx}1/\gls{rx}2 boresight                                  & $0^\circ$/$240^\circ$/$120^\circ$                  \\
            Number of \gls{tx}/\gls{rx} antennas $N$                                & 32                                                 \\
            Default transmit power $P_{tx}$                                         & 52 dBm                                             \\
            Carrier frequency $f_c$                                                 & 2.6 GHz                                            \\
            Inter-element spacing $d$                                               & $ {c}/{2f_c} \approx 5.77$ cm                      \\
            Number of subcarriers $N_c$                                             & $12\times 273=3276$                                \\
            Subcarrier spacing $\Delta f$                                           & 30 kHz                                             \\
            \gls{ifft} sub-band width $M_{\mathrm{sub}}$                            & 101                                                \\
            Range \gls{ifft} size $N_{\mathrm{IFFT}}$                               & $16{,}384$                                         \\
            Target \gls{rcs} $\sigma_{\text{rcs}}$                                  & 0.1 m\textsuperscript{2}                           \\
            Noise PSD $N_0$                                                         & $-174$ dBm/Hz                                      \\
            Rainbow beam coverage $[\theta_{\mathrm{start}},\theta_{\mathrm{end}}]$ & $[-60^\circ,\, 60^\circ]$                          \\
            \hline
        \end{tabular}%
    }
    \vspace{-1em}
\end{table}

\subsection{Receiver Architecture Comparison}
\label{subsec:rx_architecture}

We evaluate the three receiver architectures under identical simulation parameters and system settings. The \gls{fd} receiver exploits $N$ \gls{rf} chains for joint angle--range estimation via a 2D \gls{fft}. Although both \gls{pa} and \gls{pta} operate with a single \gls{rf} chain, \gls{pa} can only resolve the bistatic range within a single \gls{ofdm} symbol. The target position is therefore obtained from the range-only \gls{ls} problem
\begin{equation}
    \hat{\mathbf{p}}_{\scriptscriptstyle\mathrm{PA}} = \operatorname*{arg\,min}_{\mathbf{p}} \sum_{i=1}^{2}\bigl(\hat{d}_i - d_i(\mathbf{p})\bigr)^2,
    \label{eq:pa_ls}
\end{equation}
initialized by a coarse grid search over the \gls{roi} and refined via Nelder--Mead optimization. Both \gls{fd} and \gls{pta} recover angle and range simultaneously, and therefore adopt the GDOP-WLS fusion method in~\eqref{eq:wls_cost}.

\begin{table}[!t]
    \centering
    \caption{Receiver Architecture Comparison}
    \label{tab:rx_architecture}
    \vspace{-0.5em}
    \begin{tabular}{l c c c c}
        \hline
        \multirow{2}{*}{\textbf{Architecture}} & \multicolumn{3}{c}{\textbf{\Gls{rmse}}} & \multirow{2}{*}{\textbf{Latency (ms)}}                              \\
        \cline{2-4}
        \rule{0pt}{2.4ex}                      & \textbf{\gls{aoa} ($^\circ$)}           & \textbf{Range (m)}                     & \textbf{Pos.\ (m)} &       \\
        \hline
        \gls{fd}                               & 0.08                                    & 0.34                                   & 0.33               & 62.82 \\
        \gls{pta}                              & 0.54                                    & 1.71                                   & 1.75               & 2.26  \\
        \gls{pa}                               & ---                                     & 0.33                                   & 31.77              & 6.56  \\
        \hline
    \end{tabular}
    \vspace{-1.5em}
\end{table}

As shown in Table~\ref{tab:rx_architecture}, \gls{fd} achieves the lowest position \gls{rmse} (0.33~m) by leveraging the full $N$-element spatial aperture, at the expense of $N$ \gls{rf} chains and substantially higher computational latency. Although \gls{pa} attains comparable per-link range accuracy with a single \gls{rf} chain, the absence of angular measurements limits its fusion to a two-ellipse intersection~\eqref{eq:pa_ls} that is inherently sensitive to range estimation errors, yielding a position \gls{rmse} of 31.77~m. By recovering both angle and range from a single \gls{rf} chain, \gls{pta} navigates this trade-off with a position \gls{rmse} of 1.75~m and an order-of-magnitude reduction in latency relative to \gls{fd}.

\subsection{Baseline Fusion Schemes for PTA Architecture}
\label{subsec:baselines}

For performance comparison, we evaluate two analytical baselines. Both adopt the identical decentralized parameter estimation pipeline (Section~\ref{subsec:estimation}) as our proposed parameter-level schemes, differing only in their collaborative position fusion strategies. Fig.~\ref{fig:framework} summarizes the pipeline structure of all evaluated schemes.

\subsubsection{GI-PL (Geometric Initialization, Path-Loss-weighted Fusion)}
This scheme uses the collaborative fusion algorithm from~\cite{liu2024cooperative}. Unlike the $\sigma$-normalized squared residuals in~\eqref{eq:wls_cost}, it minimizes absolute distance and angle residuals weighted by path-loss terms $\alpha_i = R_{tx}^{-2} R_{rx,i}^{-2}$ and $\gamma_i = R_{tx}^{-2} R_{rx,i}^{-1}$, respectively. Consequently, the position is estimated as
    {\small
        \begin{equation}
            \hat{\mathbf{p}} = \operatorname*{arg\,min}_{\mathbf{p}} \left(\frac{\sum_{i=1}^{2} \alpha_i \bigl|\hat{d}_i - d_i(\mathbf{p})\bigr|}{\sum_{i=1}^2 \alpha_i}   + \frac{\sum_{i=1}^{2} \gamma_i \bigl|\hat{\theta}_i \ominus \theta_i(\mathbf{p})\bigr|}{\sum_{i=1}^2 \gamma_i}\right).
            \label{eq:dipl_cost}
        \end{equation}
    }

To initialize the iterative solver, a direct ray-based geometric intersection is applied without geometry-dependent weighting. The resulting initial position estimate $\mathbf{p}_{0} = (\hat{x}_{\mathrm{init}}, \hat{y}_{\mathrm{init}})$ is evaluated as
\begin{equation}
    \begin{aligned}
        \hat{x}_{\mathrm{init}} & = \frac{y_{rx,2} - y_{rx,1} + x_{rx,1}\tan\hat{\theta}_1 - x_{rx,2}\tan\hat{\theta}_2}{\tan\hat{\theta}_1 - \tan\hat{\theta}_2}, \\
        \hat{y}_{\mathrm{init}} & = \frac{x_{rx,2} - x_{rx,1} + y_{rx,1}\cot\hat{\theta}_1 - y_{rx,2}\cot\hat{\theta}_2}{\cot\hat{\theta}_1 - \cot\hat{\theta}_2}.
    \end{aligned}
\end{equation}

\subsubsection{GDOP-PL (GDOP Initialization, Path-Loss-weighted Fusion)}
GDOP-PL uses the same \gls{gdop}-weighted initialization as GDOP-WLS~\eqref{eq:weight_init}, but retains the path-loss-weighted cost function~\eqref{eq:dipl_cost} for iterative refinement.

\subsection{Performance and Complexity Evaluation}
\label{subsec:performance}

\subsubsection{Spatial Error Analysis}

Fig.~\ref{fig:heatmap} illustrates the spatial distribution of the mean positioning error within the hexagonal \gls{roi}. Among the baselines, GI-PL exhibits a distinct high-error band along the line connecting \gls{rx}1 and \gls{rx}2, whereas GDOP-PL eliminates this band but introduces accuracy degradation along the radial trajectories from the \gls{tx} to each \gls{rx}.

The proposed parameter-level schemes, GDOP-WLS and PF-MLP, yield a smoother error distribution, suppressing the localized error surges caused by geometric singularities. The signal-level scheme SF-CNN achieves uniformly low errors across the entire region.

\subsubsection{Accuracy, Complexity, and Robustness}

\looseness=-1
Table~\ref{tab:performance} summarizes the positioning accuracy and computational cost. The geometric vulnerabilities of GI-PL cause severe outlier amplification, inflating the \gls{rmse} to 11.71~m despite a comparable 90th-percentile error to GDOP-PL (Fig.~\ref{fig:cdf}). Replacing its geometric initialization with the proposed \gls{gdop}-weighted initialization (GDOP-PL) reduces the \gls{rmse} from 11.71~m to 3.72~m under the same objective, yet its residuals remain weighted by path-loss terms rather than by the measurement error statistics. The proposed GDOP-WLS and PF-MLP suppress the error variance more effectively, with the 90th-percentile error dropping from 3.80~m to 1.85~m and 1.87~m, lowering the \gls{rmse} to 1.75~m and 1.66~m with negligible latency overhead. SF-CNN achieves the lowest \gls{rmse} of 0.44~m; its 7.4~M-parameter model incurs the highest CPU latency (8.88~ms), which is reduced to 1.35~ms under GPU acceleration.

\begin{table}[!t]
    \centering
    \caption{Positioning Performance and Computational Cost}
    \label{tab:performance}
    \vspace{-0.5em}
    \resizebox{\columnwidth}{!}{%
        \begin{tabular}{l c c c c c c}
            \hline
            \multirow{2}{*}{\textbf{Scheme}} & \multicolumn{2}{c}{\textbf{Positioning Error (m)}} & \multicolumn{2}{c}{\textbf{Complexity}} & \multicolumn{2}{c}{\textbf{Latency (ms)}}                                                \\
            \cline{2-7}
            \rule{0pt}{2.4ex}                & \textbf{\gls{rmse}}                  & \textbf{90th Percentile}            & \textbf{\#Params}                         & \textbf{FLOPs} & \textbf{CPU} & \textbf{GPU} \\
            \hline
            GI-PL                            & 11.71                                    & 3.80                                    & ---                                       & ---            & 4.11         & ---          \\
            GDOP-PL                          & 3.72                                     & 3.70                                    & ---                                       & ---            & 3.71         & ---          \\
            GDOP-WLS                         & 1.75                                     & 1.85                                    & ---                                       & ---            & 2.26         & ---          \\
            PF-MLP                           & 1.66                                     & 1.87                                    & 9.0~k                                     & 10.0~k         & 1.74         & 1.52         \\
            SF-CNN                           & 0.44                                     & 0.64                                    & 7.4~M                                     & 313~M          & 8.88         & 1.35         \\
            \hline
        \end{tabular}%
    }
    \vspace{-1.5em}
\end{table}

Fig.~\ref{fig:rmse_vs_ptx} compares the position \gls{rmse} of both generalized and power-matched model variants for $P_{tx}$ from 42 to 52~dBm. All schemes degrade at low transmit power, with the analytical baselines reaching 40--54~m \gls{rmse} at 42~dBm. Power-matched training maintains SF-CNN at near-constant \gls{rmse} across the entire range and substantially improves PF-MLP in the low-power regime. From a latency perspective, the analytical and parameter-level schemes all operate within 4.11~ms on CPU, whereas SF-CNN requires GPU acceleration to bring its latency from 8.88~ms down to 1.35~ms, achieving the lowest \gls{rmse} at an inference speed comparable to the lightweight alternatives.

\section{Conclusion}
\label{sec:conclusion}

\looseness=-1
This paper presents a \gls{pta}-based multistatic sensing framework for \gls{uav} localization that integrates rainbow beamforming with multi-level collaborative fusion. The \gls{pta} receiver recovers both angle and range from a single \gls{rf} chain, offering a favorable accuracy--complexity trade-off between the \gls{pa} and \gls{fd} receivers at an order-of-magnitude lower latency than \gls{fd}. Two parameter-level fusion schemes, GDOP-WLS and PF-MLP, suppress geometry-induced error hotspots with negligible computational overhead. A signal-level scheme, SF-CNN, bypasses cascaded parameter estimation and attains sub-meter positioning accuracy; its higher model complexity is effectively offset by GPU acceleration. Power-sweep analysis further shows that power-matched training maintains the neural-network-based schemes' accuracy across a wide power range. These results suggest that the proposed framework offers flexible operating points along the accuracy--complexity frontier, enabling the receiver architecture and fusion strategy to be selected according to application-specific accuracy, latency, and hardware constraints.

\looseness=-1
Future work will extend the proposed framework to three-dimensional localization with elevation estimation, as well as multi-target scenarios and larger-scale networks with multiple transmitters. Additionally, practical impairments such as \gls{ttd} quantization errors and inter-node synchronization offsets will be incorporated, and topology generalization of the neural-network-based schemes via transfer learning will be investigated.

\bibliographystyle{IEEEtran}
\bibliography{reference}

\end{document}